\documentclass[prd,preprint,amsmath,amssymb,nofootinbib,eqsecnum]{revtex4-1}
\usepackage{latexsym,graphics,color,epsfig,hyperref,mathrsfs}

\newcommand{\ie}{{i.e.}}

\newcommand{\viz}{{viz.}}
\newcommand{\wrt}{with respect to}
\newcommand{\lhs}{left-hand side}

\newcommand{\be}{\begin{equation}}
\newcommand{\ee}{\end{equation}}
\newcommand{\bse}{\begin{subequations}}
\newcommand{\ese}{\end{subequations}}

\newcommand{\bcf}{\begin{center}\begin{figure}}
\newcommand{\ecf}{\end{figure}\end{center}}

\newcommand{\bct}{\begin{center}\begin{table}}
\newcommand{\ect}{\end{table}\end{center}}

\newcommand{\eq}[1]{(\ref{eq:#1})}

\newcommand{\eqs}[2]{(\ref{eq:#1}) and~(\ref{eq:#2})}
\newcommand{\eqss}[3]{(\ref{eq:#1}), (\ref{eq:#2}) and~(\ref{eq:#3})}

\newcommand{\sect}[1]{section~\ref{sec:#1}}

\newcommand{\D}{d}
\newcommand{\Int}[1]{\int \!\! d^{\D} \! #1 \,}

\newcommand{\DD}[1]{\delta^{(#1)}}

\newcommand{\der}[2]{\frac{d #1}{d #2}}

\newcommand{\pder}[2]{\frac{\partial #1}{\partial #2}}

\newcommand{\fder}[2]{\frac{\delta #1}{\delta #2}}

\newcommand{\dfder}[3]{\frac{\delta^2 #1}{\delta #2 \, \delta #3}}

\newcommand{\partialprod}[3]{D_{\underline{#1}_{#2}^{#3}}}

\newcommand{\Or}{\mathrm{O}}
\newcommand{\order}[1]{\Or \bigl( #1 \bigr)}

\newcommand{\hf}{\frac{1}{2}}

\newcommand{\one}{1\!\mathrm{l}}

\newcommand{\Tr}{\mathrm{Tr}\,}

\newcommand{\dil}[1]{D^{(#1)}}
\newcommand{\sct}[2]{{K^{(#1)}}_{#2}}

\newcommand{\dilL}[1]{\overleftarrow{D}^{(#1)}}
\newcommand{\sctL}[2]{{\overleftarrow{K}^{(#1)}}_{#2}}

\newcommand{\acom}[2]{\bigl\{#1,#2\bigr\}}
\newcommand{\comm}[2]{\bigl[#1,#2\bigr]}

\newcommand{\dfield}{\varphi}
\newcommand{\efield}{\Phi}

\newcommand{\cutoff}{K}
\newcommand{\regulator}{R}
\newcommand{\ep}{\mathcal{G}}

\newcommand{\Stot}{S}
\newcommand{\Sint}{\mathcal{S}}
\newcommand{\Gint}{\Gamma}

\newcommand{\R}{\sigma^{-1}}
\newcommand{\Rinv}{\sigma}
\newcommand{\Ltot}{\hat{L}}
\newcommand{\EAL}{\hat{\gamma}}

\newcommand{\classical}{\mathcal{C}}

\newcommand{\hepth}[1]{hep-th/#1}
\newcommand{\hepph}[1]{hep-ph/#1}

\newcommand{\condmat}[1]{cond-mat/#1}

\begin{document}

\title{A Conformal Fixed-Point Equation for the Effective Average Action}
\author{Oliver J.~Rosten}
\email{oliver.rosten@gmail.com}
\affiliation{Unaffiliated}

\begin{abstract}	
\vspace{2ex}
	A Legendre transform of the recently discovered conformal fixed-point equation is constructed, providing an unintegrated equation encoding full conformal invariance within the framework of the effective average action.
\end{abstract}

\maketitle

\section{Introduction}

In~\cite{WEMT}, an equation was derived which is, in a sense, an unintegrated version of Wilson's Exact Renormalization Group (ERG) equation. However, whereas fixed-point solutions of the latter are not necessarily conformally invariant, solutions to the former enjoy full conformal symmetry. As such, it might be hoped that this new `conformal fixed-point equation' may provide a useful tool for addressing the question as to the spectrum of (local) Conformal Field Theories (CFTs).

ERG equations are phrased in terms of the Wilsonian effective action, $\Stot$. Nevertheless, for many practical applications it has proven profitable to work instead with the Legendre transform, $\Gamma$, generally referred to as the effective average action (for reviews, see~\cite{JMP-Review,Gies-Rev,Delamotte-Rev,Fundamentals,Reuter+Saueressig-Review}).
In this paper we perform the Legendre transform of the conformal fixed-point equation, to provide an equivalent version phrased in terms of what might reasonably be called an effective average Lagrangian: \ie\ a quasi-local%
\footnote{A quasi-local functional is one which exhibits a derivative expansion.}
object which integrates to $\Gamma$.

After setting conventions in \sect{Conventions}, \sect{ERG} recalls the standard Legendre transform machinery and illustrates it by transforming the canonical ERG equation of~\cite{Ball}. Following this, \sect{SCT} derives the Legendre transform of the special conformal partner of this ERG equation. From here, it is a trivial matter to deduce the form of the conformal fixed-point equation within the framework of the effective average action which is done in \sect{CFPE}, prior to concluding.

\section{Conventions}
\label{sec:Conventions}

Throughout this paper we consider theories of a single scalar field, $\dfield$, formulated in $\D$-dimensional Euclidean space. The generators of the conformal algebra represent translations, rotations, dilatations and special conformal transformations. Interpreting $\Delta$ as a scaling dimension, a Euclidean space representation involves the differential operators $\{\partial_\mu, x_\mu \partial_\nu  - x_\nu \partial_\mu, \dil{\Delta}, \sct{\Delta}{\mu}\}$, where
\be
	 \dil{\Delta} =  x\cdot \partial + \Delta,
	\qquad
	 \sct{\Delta}{\mu} = 2x_\mu \bigl(x\cdot \partial + \Delta \bigr) -x^2 \partial_\mu
.
\ee
It will be useful to define a right action of these operators. For a function of one argument the right action is just the same as the left action \viz\ $\dil{\Delta} \dfield = \dfield \dilL{\Delta}$. For a function of two arguments, the left/right action is understood to act on the first/second argument:
\be
	\dil{\Delta} F(x,y) = \bigl(x\cdot\partial_x + \Delta \bigr) F(x,y),
	\qquad
	F(x,y) \dilL{\Delta} = \bigl(y\cdot\partial_y + \Delta \bigr) F(x,y)
.
\ee
When sandwiched between functions, a dot is used to denote an integral over the shared coordinate:
\be
	\bigl(
		\dfield \cdot F
	\bigr)(y) = \Int{x} \dfield(x) F(x,y)
.
\ee
Under an integral, $\dil{\Delta}$ and $\sct{\Delta}{\mu}$ can be transferred from one object to another using integration by parts:
\be
	\dil{\Delta} \dfield \cdot F = -\dfield \cdot \dil{\D-\Delta} F,
	\qquad
	\sct{\Delta}{\mu} \dfield \cdot F = -\dfield \cdot \sct{\D-\Delta}{\mu} F.
\label{eq:IntByParts}
\ee

Though most of the analysis of this paper will be phrased in position space, at certain junctures things will be more transparent in momentum space; upon transferring to the latter,
\be
	2 \Delta + x\cdot \partial_x + y \cdot \partial_y
	\rightarrow
	2 \Delta - \D - p \cdot \partial_p
.
\label{eq:dil-MomSpace}
\ee
We will generally overload notation so that the same symbol is used for both a function of position and its Fourier transform.

\section{Legendre Transform of the ERG}
\label{sec:ERG}

An equation for the effective average action which corresponds to the Legendre transform of the Polchinski equation~\cite{pol} was developed in~\cite{Bonini-1PI,Wetterich-1PI,Ellwanger-1PI,TRM-ApproxSolns}. However, in this paper we will not start from Polchinski's equation, but rather the canonical ERG equation of Ball et al.~\cite{Ball} which is phrased in dimensionless variables and, as such, is more appropriate for a discussion of conformal symmetry.  Since conformal symmetry is at the heart of this paper, we write ERG and associated equations directly in their fixed-point form.

\subsection{Set-up}

To define an ERG equation requires a quasi-local, ultraviolet cutoff function, $\cutoff(x,y)$. In momentum space, quasi-locality amounts to $\cutoff(p^2)$ having a Taylor expansion for small $p^2$. To be a good cutoff function, $\cutoff(p^2)$ dies off rapidly for large momentum.
From $\cutoff$ it is useful to define two objects, $\ep$ and $G$ via:
\bse
\begin{align}
	\ep^{-1} & = -\partial^2 \cutoff^{-1},
\label{eq:InvProp}
\\
	\dil{\D/2} \cutoff + \cutoff \dilL{\D/2} & = \partial^2 G
.
\label{eq:G}
\end{align}
\ese
Notice that $\ep^{-1}$ has the form of a regularized kinetic term whereas, recalling~\eq{dil-MomSpace}, $G$ is the momentum-space derivative of the cutoff function: 
\be
	p \cdot \partial_p \cutoff(p^2) = p^2 G(p^2)
.
\label{eq:G-mom}
\ee

The canonical ERG equation, here phrased in terms of the full Wilsonian effective action%
\footnote{In~\cite{Ball}, the equation is written in terms of $\Sint = \Stot - \hf \dfield \cdot \ep^{-1} \cdot \dfield$.},
 takes the form:
\be
	\dil{\delta} \dfield \cdot \fder{\Stot[\dfield]}{\dfield}
	+
	\dfield \cdot G \cdot \ep^{-1} \cdot \fder{\Stot}{\dfield}
	-
	\hf
	\fder{\Stot}{\dfield} \cdot G \cdot \fder{\Stot}{\dfield}
	+
	\hf
	\fder{}{\dfield} \cdot G \cdot \fder{\Stot}{\dfield}
	=
	0
,
\label{eq:ERGE}
\ee
where $\delta$ is the scaling dimension of $\dfield$, frequently expressed in terms of the quantity, $\eta$:
\be
	\delta = \frac{\D-2+\eta}{2}
.
\label{eq:delta}
\ee
The Legendre transform was constructed in~\cite{HO-Remarks} (see also~\cite{OJR-AsymptoticSafety}). We will re-derive it in the next section, as a warm-up for subsequent calculations.

To complete the set-up, let us define the object, $\R$ which, in momentum space, is given by
\be
	\R(p^2) = p^{2(\eta/2-1)} \cutoff(p^2)
	\int_0^{p^2} d q^2
	q^{-2(\eta/2)}
	\der{}{q^2} 
	\frac{1}{\cutoff(q^2)}
.
\label{eq:rho}
\ee
Utilizing~\eq{dil-MomSpace}, it is easy to confirm that
\be
	\dil{\delta} \cutoff^{-1} \cdot \R + \cutoff^{-1} \cdot \R \dilL{\delta}
	=
	\cutoff^{-1} \cdot G \cdot \cutoff^{-1}
\label{eq:dR}
.
\ee

The effective average action, $\Gint[\efield]$, is defined according to~\cite{HO-Remarks}
\be
	\Gint[\efield] = \Stot[\dfield] 
	-
	\hf
	\bigl(
		\dfield - \efield \cdot \cutoff
	\bigr)
	\cdot
	\cutoff^{-1} \cdot \Rinv
	\cdot
	\bigl(
		\dfield - \cutoff\cdot \efield
	\bigr)
,
\ee
with $\efield$ defined via:
\be
	\fder{\Stot[\dfield]}{\dfield}
	=
	\Rinv
	\cdot
	\bigl(
		\cutoff^{-1} \cdot \dfield  - \efield
	\bigr)
.
\label{eq:dS/dphi}
\ee
Note that $\Rinv \cdot \cutoff^{-1} = \cutoff^{-1} \cdot \Rinv$ (as is readily seen in momentum space, since both functions just depend on $p^2$). 
The equation complimentary to~\eq{dS/dphi} reads:
\be
	\fder{\Gint[\efield]}{\efield}
	=
	\Rinv
	\cdot
	\bigl(
		\dfield - \cutoff \cdot \efield
	\bigr)
,
\label{eq:dG/dPhi}
\ee
implying the useful result
\be
	\fder{\Stot[\dfield]}{\dfield}
	=
	\cutoff^{-1}
	\cdot
	\fder{\Gint[\efield]}{\efield}
.
\label{eq:fder-Relationship}
\ee

\subsection{Derivation}

Focussing on the first term of the ERG equation~\eq{ERGE}, we substitute for $\dfield$ using~\eq{dG/dPhi}, for $\delta S / \delta \dfield$ using~\eq{fder-Relationship} and utilize~\eq{IntByParts}
to deduce that
\be
	\dil{\delta} \dfield \cdot \fder{\Stot[\dfield]}{\dfield}
	=
	-\biggl(
		\efield \cdot \cutoff 
		+\fder{\Gint}{\efield} \cdot \R 
	\biggr)
	\cdot \dil{\D-\delta} \cutoff^{-1} \cdot \fder{\Gint}{\efield}
.
\label{eq:firstTerm}
\ee
To process the first term we integrate by parts and then re-express
\be
	\efield \cdot \cutoff \dilL{\delta} \cdot \cutoff^{-1}
	=
	\dil{\D-\delta} \efield 
	+
	\efield \cdot 
	\bigl(
		\dil{\delta} \cutoff + \cutoff \dilL{\delta} 
	\bigr)
	\cdot \cutoff^{-1}
.
\label{eq:FirstTerm-Process}
\ee
From~\eq{G},
\be
	\dil{\delta} \cutoff + \cutoff \dilL{\delta} = 
	(2\delta - \D) \cutoff
	+
	\partial^2 G
\label{eq:Ddelta-K}
\ee
and so, noting from~\eq{InvProp} that $\partial^2 G \cdot \cutoff^{-1} = -G\cdot \ep^{-1}$ (which, again, is obvious in momentum space):
\be
	\efield \cdot \cutoff \dilL{\delta} \cdot \cutoff^{-1} \cdot \fder{\Gint}{\efield}
	=
	\dil{\delta} \efield \cdot \fder{\Gint}{\efield}	
	-
	\efield \cdot G \cdot \ep^{-1} \cdot \fder{\Gint}{\efield}
.
\ee
Performing similar manipulations on the second term in~\eq{firstTerm}, but this time exploiting~\eq{dR} together with
\be
	\bigl(
		\dil{\D-\delta} \cutoff^{-1} + \cutoff^{-1} \dilL{\D-\delta}
	\bigr)
	=
	(d-2\delta) \cutoff^{-1} + \cutoff^{-1} \cdot G \cdot \ep^{-1}
\label{eq:D-Kinv}
\ee
yields
\be
	\dil{\delta} \dfield \cdot \fder{\Stot}{\dfield}
	= 
	\dil{\delta} \efield \cdot \fder{\Gint}{\efield}	
	-
	\biggl(
		\efield 
		+\fder{\Gint}{\efield} \cdot \R \cdot \cutoff^{-1}
	\biggr)
	\cdot G \cdot \ep^{-1} \cdot \fder{\Gint}{\efield}
	- \fder{\Gint}{\efield} \cdot \R \cdot \cutoff^{-1} \cdot \dil{\D-\delta} \fder{\Gint}{\efield}
.
\label{eq:dilS}
\ee
The terms enclosed in the big brackets combine to give $\dfield \cdot \cutoff^{-1}$, according to~\eq{dG/dPhi}. Taking advantage of the fact that we can move $\cutoff^{-1}$ to the right \viz\ $\cutoff^{-1} \cdot G \cdot \ep^{-1} =  G \cdot \ep^{-1} \cdot \cutoff^{-1}$, as is particularly clear in momentum space, makes it manifest that we can utilize~\eq{fder-Relationship} to give
\[
	-\dfield \cdot G \cdot \ep^{-1} \cdot \fder{\Stot}{\dfield}
.
\]
Returning to~\eq{dilS} and utilizing the fact that $\R \cdot \cutoff^{-1} = \cutoff^{-1} \cdot \R$, the final term can be processed to give:
\be
	\hf
	\fder{\Gint}{\efield} \cdot 
	\bigl(
		\dil{\delta} \R \cdot \cutoff^{-1}  + \R \cdot \cutoff^{-1} \dilL{\delta}
	\bigr)
	\cdot
	\fder{\Gint}{\efield}
	=
	\hf
	\fder{\Gint}{\efield} \cdot 
	\cutoff^{-1} \cdot G \cdot \cutoff^{-1} \cdot
	\fder{\Gint}{\efield}
	= 
	\hf
	\fder{\Stot}{\dfield} \cdot G \cdot \fder{\Stot}{\dfield}
\ee
where, in the first step, we have used~\eq{dR} and in the second~\eq{fder-Relationship}. Therefore,
\be
	\dil{\delta} \dfield \cdot \fder{\Stot[\dfield]}{\dfield}
	+
	\dfield \cdot G \cdot \ep^{-1} \cdot \fder{\Stot}{\dfield}
	-
	\hf
	\fder{\Stot}{\dfield} \cdot G \cdot \fder{\Stot}{\dfield}
	=
	\dil{\delta} \efield \cdot \fder{\Gint[\efield]}{\efield}
,
\ee
leaving just the final term of the ERG equation~\eq{ERGE} to transform.

Differentiating~\eq{dS/dphi} \wrt\ $\dfield$ and~\eq{dG/dPhi} \wrt\ $\efield$ gives:
\bse
\begin{align}
	\R \cdot \fder{}{\dfield} \fder{\Stot}{\dfield}
	& =
	-\fder{\efield}{\dfield} + \cutoff^{-1}
\label{eq:d^2S/dphidphi}
\\
	\R \cdot \fder{}{\efield} \fder{\Gint}{\efield}
	& =
	\fder{\dfield}{\efield} - \cutoff
,
\label{eq:d^2G/dPhidPhi}
\end{align}
\ese
where we understand, for some $A,B$
\be
	\biggl(\fder{A}{B}\biggr)(x,y) = \fder{A(x)}{B(y)},
	\qquad
	\biggl(\dfder{}{A}{B}\biggr)(x,y) = \fder{}{A(x)} \fder{}{B(y)}
.
\ee
Equations~\eqs{d^2S/dphidphi}{d^2G/dPhidPhi} can be combined to give
\be
	\dfder{\Stot}{\dfield}{\dfield}
	=
	- \Rinv \cdot
	\biggl(
		\dfder{\Gint}{\efield}{\efield} + \Rinv \cdot \cutoff
	\biggr)^{-1} \cdot \Rinv
	+ \Rinv \cdot \cutoff^{-1}
.
\label{eq:ddS}
\ee
Finally, then, up to a divergent vacuum term which, formally, may be written $-\hf \Tr G \cdot \Rinv \cdot \cutoff^{-1}$
\be
	\dil{\delta} \efield \cdot \fder{\Gint[\efield]}{\efield}
	=
	\hf \Tr
	\Bigl[
		\Rinv \cdot G \cdot \Rinv \cdot
		\bigl(
			\Gint^{(2)} + \Rinv \cdot \cutoff
		\bigr)^{-1}
	\Bigr]
\label{eq:ERG-LT}
\ee
where, as usual, we define $\Gint^{(2)}$ to be the second functional derivative of $\Gamma$.

This is of a similar general form to the more familiar equation for the effective average action found in the literature; the differences arise because~\eq{ERG-LT} is the Legendre transform of the canonical ERG equation, \eq{ERGE}, rather than Polchinski's equation. Note that, as can be readily checked using~\eq{rho}, $\Rinv \cdot \cutoff$ behaves like an infrared regulator, remaining non-zero for vanishing momentum and vanishing for large momentum. Indeed, as pointed out in~\cite{HO-Remarks}, if we define
\be
	\regulator = \Rinv \cdot \cutoff
\label{eq:regulator}
\ee
then, from~\eq{dR},
\be
	\dil{\D-\delta} \regulator + \regulator \dilL{\D-\delta} = -\Rinv \cdot G \cdot \Rinv.
\label{eq:regulator-relationship}
\ee
In momentum space this becomes
\be
	\bigl(
		p \cdot \partial_p - 2 + \eta
	\bigr)
	\regulator(p^2)
	= 
	G(p^2) \Rinv^2(p^2)
,
\label{eq:regulator-relationship-mom}
\ee
as can verified using~\eqss{G-mom}{rho}{regulator}.

It makes sense to reinterpret $\regulator$ as an independent infrared regulator and so we rewrite~\eq{ERG-LT} as~\cite{HO-Remarks}
\be
	\dil{\delta} \efield \cdot \fder{\Gint[\efield]}{\efield}
	=
	\hf \Tr
	\Bigl[
		\bigl(\dot{\regulator} + \eta \regulator \bigr)
		\bigl(
			\Gint^{(2)} + \regulator
		\bigr)^{-1}
	\Bigr]
,
\label{eq:ERG-LT-Canonical}
\ee
with the understanding that all $\eta$-dependence is now explicit and
\be
	-\dot{\regulator} \equiv \dil{\D/2+1} \regulator + \regulator \dilL{\D/2+1} 
	\qquad
	\mathrm{or}
	\qquad
	\dot{\regulator}(p^2) 
	= \bigl( p\cdot \partial_p - 2 \bigr) \regulator(p^2)
.
\label{eq:regulator_dot}
\ee

\section{Special Conformal Transformations}
\label{sec:SCT}

Just as the ERG equation~\eq{ERGE} encodes dilatation invariance, so too is there an equation encoding special conformal invariance~\cite{Representations,WEMT}:
\be
	\sct{\delta}{\mu} \dfield \cdot \fder{\Stot}{\dfield}
	+
	\dfield \cdot \ep^{-1} \cdot G_\mu \cdot \fder{\Stot}{\dfield}
	- \eta \partial_\mu \dfield \cdot \cutoff^{-1} \cdot G \cdot \fder{\Stot}{\dfield}
	-
	\hf
	\fder{\Stot}{\dfield} \cdot G_\mu \cdot \fder{\Stot}{\dfield}
	+\hf
	\fder{}{\dfield} \cdot G_\mu \cdot \fder{\Stot}{\dfield}
	= 0
,
\label{eq:SCE}
\ee
where
\be
	G_\mu(x,y)
	\equiv 
	(x+y)_\mu G\bigl((x-y)^2\bigr)
\ee
or, equivalently but slightly more abstractly,
\be
	G_\mu = \acom{G}{X_\mu}
.
\label{eq:G_mu}
\ee

Construction of the Legendre transform proceeds similarly to before. Considering the first term in~\eq{SCE}, this can be directly re-expressed along the lines of~\eq{firstTerm}:
\be
	\sct{\delta}{\mu} \dfield \cdot \fder{\Stot[\dfield]}{\dfield}
	=
	-
	\biggl(	
		\efield \cdot \cutoff \cdot 
		+\fder{\Gint}{\efield} \cdot \R
	\biggr)		
	\cdot \sct{\D-\delta}{\mu} \cutoff^{-1} 
	\cdot \fder{\Gint}{\efield}
.
\label{eq:firstTerm-SC}
\ee
The first term can be processed, analogously to~\eq{FirstTerm-Process}, to give
\be
	\efield \cdot \cutoff \sctL{\delta}{\mu} \cdot \cutoff^{-1} \cdot \fder{\Gint}{\efield}
	=
	\sct{\D-\delta}{\mu} \efield \cdot \fder{\Gint}{\efield}
	+
	\efield
	\cdot
	\bigl(
		\sct{\delta}{\mu} \cutoff + \cutoff \sctL{\delta}{\mu}
	\bigr)
	\cdot \cutoff^{-1} \cdot
	\fder{\Gint}{\efield}
.
\ee
We may continue to closely mirror the previous analysis by exploiting the following result, utilized in~\cite{Representations}. Suppose there are two functions, $U(x,y)$, $V(x,y)$ which are, respectively, Fourier transforms of $U(p^2)$, $V(p^2)$. If they are related by
\be
		\dil{\Delta} U + U \dilL{\Delta} = V
\ee
then this implies that
\be
	\sct{\Delta}{\mu} U + U \sctL{\Delta}{\mu} = \acom{V}{X_\mu}.
\label{eq:helper}
\ee
Combining~\eq{Ddelta-K} with~\eq{helper} gives:
\be
	\sct{\delta}{\mu} \cutoff + \cutoff \sctL{\delta}{\mu}
	=
	\acom{\partial^2 G + (\eta -2) \cutoff }{X_\mu}
.
\ee
To process this, we will use the result, for some $U$ as above
\be
	\bigl[
		X_\mu, U
	\bigr]
	= 2 \partial_\mu U'
,
\label{eq:X_muU}
\ee
where, in momentum space, $U' = d U(p^2) /dp^2$. Using this result, and noting that partial derivatives acting to the right strike only the first coordinate of some $F(x,y)$,
\bse
\begin{align}
	\acom{\cutoff}{X_\mu}
	&= 2 X_\mu \cutoff - \bigl[X_\mu, \cutoff \bigr]
	= 2 X_\mu \cutoff  - \partial_\mu G
\\
	\acom{\partial^2 G}{X_\mu}
	& =
	\partial^2 \acom{G}{X_\mu} - \bigl[\partial^2, X_\mu\bigr] G
	=
	\partial^2 G_\mu - 2\partial_\mu G
\end{align}
\ese
where, for $G_\mu$, we recall~\eq{G_mu}.
Therefore,
\be
	\efield \cdot \cutoff \sctL{\delta}{\mu} \cdot \cutoff^{-1} \cdot \fder{\Gint}{\efield}
	=
	\sct{\delta}{\mu} \efield \cdot \fder{\Gint}{\efield}
	-\efield\cdot \partial^2 \cdot G_\mu \cdot \cutoff^{-1} \cdot \fder{\Gint}{\efield}
	+ \eta \, \partial_\mu \efield \cdot G \cdot \cutoff^{-1} \cdot \fder{\Gint}{\efield}
,
\label{eq:sct-1}
\ee
having integrated by parts in the final term.

Now we return to~\eq{firstTerm-SC} and treat the final term. Utilizing~\eqs{D-Kinv}{helper} gives:
\begin{multline}
	-\fder{\Gint}{\efield} \cdot \R \cdot \sct{\D-\delta}{\mu} \cutoff^{-1} \cdot \fder{\Gint}{\efield}
	=
	-
	\fder{\Gint}{\efield} \cdot \R \cdot \cutoff^{-1} \sct{\delta}{\mu} \fder{\Gint}{\efield}
\\
	-\fder{\Gint}{\efield} \cdot \R \cdot
	\bigl\{
		(2-\eta) \cutoff^{-1} + \cutoff^{-1} \cdot G \cdot \ep^{-1}, X_\mu
	\bigr\}
	\cdot \fder{\Gint}{\efield}	
.
\end{multline}
To process this, first note that $\cutoff^{-1} \cdot G \cdot  \ep^{-1} =\ep^{-1} \cdot G \cdot \cutoff^{-1}$. Now expanding out the last term, we can move the $X_\mu$ around with impunity: commutators with $X_\mu$ give rise to $\partial_\mu$ terms which here vanish due to the resulting integrands being odd. 
Therefore,
\begin{align}
	-\fder{\Gint}{\efield} \cdot \R \cdot \sct{\D-\delta}{\mu} \cutoff^{-1} \cdot \fder{\Gint}{\efield}
	&=
	\hf \fder{\Gint}{\efield} \cdot 
	\bigl(
		\sct{\delta}{\mu} \R \cdot \cutoff^{-1} +  \R \cdot \cutoff^{-1} \sctL{\delta}{\mu}
	\bigr)	
	 \fder{\Gint}{\efield}
\nonumber
\\
	&
	\qquad
	 -
	 \fder{\Gint}{\efield}
	 \cdot
	 \R
	 \cdot
	 \ep^{-1} \cdot G_\mu \cdot \cutoff^{-1}
	 \cdot
	 \fder{\Gint}{\efield}
\nonumber
\\
	&
	=
	\hf
	\fder{\Gint}{\efield}
	 \cdot
	\bigl(
		 \cutoff^{-1} \cdot G_\mu \cdot \cutoff^{-1}
		- 2\R \cdot \ep^{-1} \cdot G_\mu \cdot \cutoff^{-1}
	\bigr)	 
	 \cdot
	 \fder{\Gint}{\efield}
\label{eq:sct-2}
\end{align}
where we have utilized~\eqs{dR}{helper} and have exploited the freedom to move $X_\mu$ around to phrase the result in terms of $G_\mu$.

This completes the treatment of the terms arising in~\eq{firstTerm-SC}, with the final results given in~\eqs{sct-1}{sct-2}. All bar the first term of the former cancel against the second, third and fourth terms in the special conformal partner of the ERG equation, \eq{SCE}. The final term of the latter equation can be treated using~\eq{ddS}, to give the desired Legendre transform:
\be
	\sct{\delta}{\mu} \efield \cdot \fder{\Gint}{\efield}
	=
	\hf
	\Tr
	\Bigl[
		\Rinv \cdot G_\mu \cdot \Rinv \cdot
		\bigl(
			\Gint^{(2)} + \Rinv \cdot \cutoff
		\bigr)^{-1}
	\Bigr]
.
\label{eq:SCE-LT}
\ee
As with the Legendre transform of the ERG equation, we can re-express this in terms of $\regulator$.  Let us start by recalling~\eq{G_mu} and observing that
\be
	\Rinv \cdot \acom{G}{X_\mu} \cdot \Rinv
	=
	\acom{\Rinv \cdot G \cdot \Rinv}{X_\mu}
	+
	\comm{\Rinv}{X_\mu}  \cdot G \cdot \Rinv
	-
	\Rinv \cdot G \cdot \comm{\Rinv}{X_\mu}
	 = \acom{\Rinv \cdot G \cdot \Rinv}{X_\mu}
,
\ee
where the cancellation may be readily justified by using~\eq{X_muU} and working in momentum space.
Recalling~\eq{ERG-LT-Canonical}, we have:
\be
	\sct{\delta}{\mu} \efield \cdot \fder{\Gint}{\efield}
	=
	\hf
	\Tr
	\Bigl[
		\acom{\dot{\regulator} + \eta \regulator}{X_\mu}
		\bigl(
			\Gint^{(2)} + \regulator
		\bigr)^{-1}
	\Bigr]
.
\label{eq:SCE-LT-Canonical}
\ee
A similar equation, derived in a more heuristic manner, was presented in~\cite{Delamotte-Conformal}.

\section{Conformal Fixed-Point Equation}
\label{sec:CFPE}

The conformal fixed-point equation of~\cite{WEMT} can be expressed as follows. First, take $\Ltot$ to be in the equivalence class of quasi-local objects which integrate to the Wilsonian effective action; next define
\be
	\partialprod{\alpha}{j}{i}
	\equiv
	\Biggl\{
	\begin{array}{cl}
		\prod_{k=j}^i \partial_{\alpha_k} & i \geq j,
	\\
		1 & i < j.
	\end{array}
\ee
Using $\times$ to indicate that the multiplied objects are at the same point we now introduce
\begin{multline}
	\classical_{\Ltot}(\dfield)
	=
	\delta \dfield \times \fder{\Stot}{\dfield}
	-\D\Ltot
	+
	\sum_{i=1}^\infty
	\bigl[ \partialprod{\sigma}{1}{i}, x\cdot\partial \bigr] \dfield
	\times
	\pder{\Ltot}{(\partialprod{\sigma}{1}{i}\dfield)}
\\
	-\partial_\lambda
	\biggl(
		\hf
		\bigl(
			\delta_{\omega \lambda} \delta_{\rho\sigma}
			- 2\delta_{\omega \rho} \delta_{\sigma\lambda}
		\bigr)
		\sum_{i=2}^\infty
		\bigl[
			\bigl[
				\partialprod{\sigma}{1}{i}, x_\sigma
			\bigr],
			x_\rho \partial_\omega
		\bigr]
		\dfield
		\times
		\pder{\Ltot}{(\partialprod{\sigma}{1}{i} \dfield)}
	\biggr)
.
\end{multline}
As demonstrated in~\cite{WEMT}---and making the coordinate dependence explicit---$ \classical_{\Ltot}(\dfield; x)$ has the following properties:
\be
	\Int{x} \classical_{\Ltot}(\dfield; x)
	=
	\dil{\delta} \dfield \cdot \fder{\Stot}{\dfield}
	,
	\qquad
	\Int{x} 2 x_\mu \classical_{\Ltot}(\dfield; x)
	=
	\sct{\delta}{\mu} \dfield \cdot \fder{\Stot}{\dfield}
.
\ee
With this in mind, and using the notation
\be
	\acom{G}{\one}(y,z;x) = G(y,x) \DD{\D}(x-z) + \DD{\D}(y-x) G(x,z)
,
\label{eq:acom}
\ee
the conformal fixed-point equation reads:
\begin{multline}
	\order{\partial^2}
	=
	\classical_{\Ltot} (\dfield)
	+
	\hf \dfield \cdot \ep^{-1} \cdot \acom{G}{\one} \cdot \fder{\Stot}{\dfield}
	-\hf \fder{\Stot}{\dfield} \cdot G \times \fder{\Stot}{\dfield}
	+\hf \fder{}{\dfield} \cdot G \times \fder{\Stot}{\dfield}
\\
	+
	\partial_\lambda
	\biggl(
		\frac{\eta}{4} \partial_\lambda \dfield \cdot \cutoff^{-1} 
		\cdot \acom{G}{\one} \cdot \fder{\Stot}{\dfield}
	\biggr)
,
\label{eq:CERG}
\end{multline}
with the understanding that, in $\D=2$, the \lhs\ must be expressible as $\partial^2 (\mathrm{scalar})$.
Confirming this equation is easy and will allow us to directly write down the Legendre transform. First, consider integrating. The total derivative term vanishes and the remaining terms combine to produce the ERG equation~\eq{ERGE}. Secondly, multiply by $2x_\mu$ and then integrate; this time it is straightforward to check that we generate the special conformal partner of the ERG equation, \eq{SCE}.

With this in mind, let us define $\EAL$ to be in the equivalence class of objects that integrate to the effective average action, $\Gamma$. Recalling~\eqs{ERG-LT}{SCE-LT}, it must be that
\be
	\order{\partial^2}
	=
	\classical_{\EAL} (\efield)
	-
	\frac{1}{4} \Tr
	\Bigl[
		\Rinv \cdot \acom{G}{\one} \cdot \Rinv \cdot
		\bigl(
			\Gint^{(2)} + \Rinv \cdot \cutoff
		\bigr)^{-1}
	\Bigr]
\label{eq:Final}
\ee
with the same restriction in $\D=2$ mentioned above. To clarify the notation, we understand the coordinate shared by $G$ and $\one$ to be unintegrated, whereas all other coordinates are integrated over. Equivalently, rewriting in terms of $\regulator$:
\be
	\order{\partial^2}
	=
	\classical_{\EAL} (\efield)
	-
	\frac{1}{4} \Tr
	\Bigl[
		\acom{\dot{\regulator} + \eta \regulator}{\one}
		\cdot
		\bigl(
			\Gint^{(2)} + \regulator
		\bigr)^{-1}
	\Bigr]
.
\label{eq:Final-Canonical}
\ee

\section{Conclusion}
\label{sec:Conclusion}

Let us summarize the main results. Starting from the fixed-point version of the canonical ERG equation, \eq{ERGE}, its Legendre transform was re-derived, giving~\eq{ERG-LT}. Turning to the special conformal partner of the ERG equation, \eq{SCE}, its Legendre transform was constructed yielding~\eq{SCE-LT}.  Just as~\eqs{ERGE}{SCE} combine to give the conformal fixed-point equation~\eq{CERG}, so do~\eqs{ERG-LT}{SCE-LT} combine to give~\eq{Final}. The latter is brought into a more friendly form in~\eq{Final-Canonical}. Structurally, \eqs{SCE}{Final} share a common contribution (modulo the replacement of $\Ltot$ by $\EAL$) but the Legendre transformed version is, overall, more compact.

It might be hoped that an advantage of \eq{Final-Canonical} compared to more standard equations for the effective average action is that conformal invariance is built in. Indeed, solutions of the ERG equation (or, equivalently, its Legendre transform) are necessarily scale invariant, but not necessarily conformally invariant. Optimistically, the additional structure present in~\eq{Final-Canonical} may allow new methods to be developed for the extraction of solutions. Quite apart from this, it should be possible to use~\eq{Final-Canonical} with conventional approximations to determine whether or not previously obtained solutions for the effective average action enjoy full conformal symmetry.

\begin{acknowledgments}
	I thank Hugh Osborn for comments on the manuscript and particularly for emphasising to me the relationships between various Legendre transformed versions of the ERG equation.
\end{acknowledgments}

\end{document}